\documentstyle[11pt,aaspp4]{article}
\tighten
 
\received{18 Aug 2000}

\slugcomment{submitted to ApJ}

\begin{document}

\title{Topology of the Galaxy Distribution in the Hubble Deep Fields}
\author{Changbom Park$^1$, J. Richard Gott III$^2$, and Y. J. Choi$^1$}
\affil{$^1$Department of Astronomy, Seoul National University,
    Seoul, 151-742 Korea;}
\affil{cbp@astro.snu.ac.kr, choiyj@astro.snu.ac.kr}
\affil{$^2$Princeton University Observatory, Peyton Hall, Princeton, NJ 08544; jrg@astro.princeton.edu}

\begin{abstract}
We have studied topology of the distribution of the high redshift galaxies 
identified in the Hubble Deep Field (HDF) North and South. 
The two-dimensional genus is measured from the projected distributions 
of the HDF galaxies at angular scales from $3.8''$ to $ 6.1''$. 
We have also divided the samples into three 
redshift slices with roughly equal number of galaxies using photometric 
redshifts to see possible evolutionary effects on the topology.

The genus curve of the HDF North clearly indicates clustering
of galaxies over the Poisson distribution while the clustering is
somewhat weaker in the HDF South. This clustering is mainly
due to the nearer galaxies in the samples.
We have also found that the genus curve of galaxies in the HDF 
is consistent with the Gaussian random phase distribution
with no significant redshift dependence.

\end{abstract}


\section{INTRODUCTION}


An important prediction of typical inflationary models is that
the matter fluctuation field has a Gaussian random phase distribution.
At large linear scales where the galaxy distribution is presumely 
still in the linear regime and therefore keeping the statistics 
of the primordial fluctuation field, 
studies of the topology of the galaxy distribution can test 
such predictions.  At smaller non-linear scales where
the galaxy distribution depends sensitively on the small-scale
physics, a study of topology can give us information on the mechanism of
galaxy formation and evolution as well as on cosmology.

During the past 15 years topology of large scale structures has
been measured by many authors using various observational samples
following after the initial work of Gott, Melott \& Dickinson (1986),
Hamilton, Gott \& Weinberg (1986), and
Gott, Weinberg \& Melott (1987).
Among these, topology measures for two-dimensional fields have been 
introduced by Coles \& Barrow (1987), Melott et al. (1989) 
and Gott et al. (1990), and
applied to observational samples like the angular distribution of galaxies
on the sky (Coles \& Plionis 1991; Gott et al. 1992), the
distribution of galaxies in slices of the universe (Park et al. 1992;
Colley 1997)
and the temperature fluctuation field of the Cosmic Microwave Background 
(Coles 1988; Gott et al. 1990; Smoot et al. 1994; Kogut et al. 1996; 
Colley, Gott \& Park  1996; Park et al. 1998). In the first two cases
where the samples dominantly include nearby galaxies,
the topology of two-dimensional galaxy distributions as revealed by
the genus or the Euler-Poincar\'e characteristic statistics, 
is consistent with the
random phase Gaussian distribution with a possible weak `meat-ball' topology.

The HDF images  (Williams et al. 1996) taken
by the Wide Field Planetary Camera 2 (WFPC-2) on the Hubble Space
Telescope, have given us an unprecedently deep view of
the high redshift universe.  One can easily detect objects with AB
magnitudes (Oke 1974) down to $28 \sim 29$ (Madau et al. 1996) and with
an angular resolution of about $0.05''$. 
Important issues one can address from the HDF data
include the discovery of proto-galaxies forming at high redshifts, or
delineating the epoch of galaxy formation (Steidel et al. 1996; 
Clements \& Couch 1996; Park \& Kim 1998).  
Another important problem one can study from the HDF data is the
properties of galaxies at high redshifts, or the evolution of galaxies.  
This can be done by looking at numbers, colors,
and clustering of galaxies as a function of redshift.

In this paper we study the topology of the distribution of
galaxies identified in
the HDF North and South and having either spectroscopic or
photometric redshifts (Fern\'andez-Soto, Lanzetta, \& Yahil 1999,
hereafter FLY99; Lanzetta et al. 1999).
From these data sets we hope to study the topology of 
the galaxy distribution at high redshifts.

\section{Topology Measure}

\subsection{The Genus}
We use the two-dimensional genus statistic introduced by Melott et 
al. (1989) as a quantitative measure of topology of the galaxy distribution
in the HDFs. The study most similar to our work is that of Gott et al.
(1992) who have measured the genus of angular distributions of
nearby galaxies in the UGC and ESO catalogs. Coles \& Plionis (1991)
have also measured the Euler-Poincar\'e characteristic, which is
equivalent to the genus in the limit of negligible sample boundary effects,
for the Lick catalog.

The two-dimensional genus is defined as (Gott et al. 1992)
$$ G = {\rm (number\; of\; isolated\; high\; density\; regions)} - 
         {\rm (number\; of\; isolated\; low\; density\; regions)}. \eqno (1)$$
When a two-dimensional distribution is given, the Gauss-Bonnett
theorem relates the genus of isodensity contours with the line integral 
of the local curvature $\kappa$ (Gott et al. 1990) 
$$ G = {1\over 2\pi}\int \kappa ds. \eqno (2) $$
In the case of a Gaussian random field the genus per unit area
is known to be (Melott et al. 1989; Coles 1988)
$$g = {1\over (2\pi)^{3/2}} 
{\langle k^2\rangle\over 2}\nu e^{-\nu^2/2},\eqno(3)$$
where $\nu$ is the threshold density for the isodensity contour
in units of standard deviations from the mean, 
$\langle k^2\rangle=\int k^2 P_2(k)d^2 k / \int P_2(k) d^2 k$  and $P_2(k)$
is the smoothed two-dimensional power spectrum.
In practice, we are not interested in the one-point distribution of the
density field, and therefore use the label $\nu_A$ which parametrizes
the area faction by
$$f_{\rm A} = {1\over \sqrt{2\pi}}\int_{\nu_A}^{\infty} e^{-t^2/2} dt. \eqno (4)$$
We calculate the genus from $\nu_A=-3$ to 3 with an interval of 0.2.

When the power spectrum of the three-dimensional galaxy distribution is
of the power-law form $P_3(k)\propto k^n$ and when the thickness of the
two-dimensional slice is much larger or smaller than the smoothing 
length of the Gaussian filter $W(r) \propto e^{-r^2/2 R_G^2}$, 
there exist simple analytic formulae for the amplitude
(Melott et al. 1989)
$$\langle k^2\rangle={F(n)\over R_G^2}, \eqno (5)$$
where $F(n) = (n+2)/2$ for $n>-2$  and $0$ for $-3<n\le -2$
for slices with thickness much larger than the smoothing length 
$R_G$, and $F(n) = 1$ for $n\ge -1$ and $(n+3)/2$
for $-3<n<-1$ for thin slices.
The thick slice approximation is relevant for maps of galaxies
projected on the sky.

\subsection{The Genus-Related statistics}

Since the genus-threshold density relation for Gaussian fields is known,
non-Gaussian behaviour of a field can be detected from deviations
from the relation.
We quantify such deviations by genus-related statistics. The first is 
the shift of the genus curve $\Delta \nu$ (Park et al. 1992). 
We measure the shift parameter from
the observed genus curve by minimizing the $\chi^2$ between the data 
and the fitting function
$$G = A \nu' e^{-\nu'^2/2}, \eqno (6)$$
where $\nu'=\nu-\Delta\nu$ and the amplitude $A$ of the genus curve
is allowed to have different values at negative and positive $\nu'$.
The $\chi^2$-minimization is performed over the range $-1.0\le \nu \le 1.0$.

The second statistic is the asymmetry parameter which measures the
difference in the amplitude of the genus curve in the positive and
negative thresholds (i.e. the difference between the numbers of clumps and
voids).
The asymmetry parameter is defined as
$$\Delta g = A_C - A_V, \eqno (7)$$
where 
$$A_C = \int_{\nu_1}^{\nu_2} g_{\rm obs} d\nu 
/ \int_{\nu_1}^{\nu_2} g_{\rm fit} d\nu, \eqno (8)$$
and likewise for $A_V$.
The integration is limited to $-2\le \nu \le -0.4$ for $A_V$ and to
$0.4\le \nu \le 2$ for $A_C$. The overall amplitude $A$ of the 
best-fit genus curve $g_{\rm fit}$
is found from the $\chi^2$-fitting over the range $-2.0\le \nu \le 2.0$.
$\Delta g$ is positive when high-density regions are divided into many
clumps while the low-density regions are merged into fewer voids.
For an observed genus curve, we therefore measure
the best-fit amplitude $A$, the shift parameter $\Delta\nu$, and
the asymmetry parameter $\Delta g$.

\section{Analysis of the HDF Data}

\subsection{The HDF data}

We use the photometric redshift data of galaxies
in the HDF North (hereafter HDFN) published by FLY99.
The catalog has 1067 galaxies with photometric and/or spectroscopic
redshifts. We use the spectroscopic redshifts whenever available.
We limit the sample to $z\le 2$, which leaves us with 820 galaxies
because the redshift space distribution of galaxies sharply
drops at $z\sim 2$ and because the photometric redshift starts to have
relatively large error at $z\ge 2$ (FLY99).
We then drop the Planetary Camera image as well as the edges of
the Wide Field Camera images where the magnitude limit to the sample
is bright, i.e. $AB(8140)=26.0$.

In Figure 1 the long-dashed lines delineate the inner part of
the WFPC2 images which encloses 714 galaxies 
with $AB(8140)\le 28.0$ (hereafter the S-zone; see dash lines 
in Fig. 1 of FLY99).
The genus is actually measured in the region 100 pixels 
inside these boundaries (hereafter the G-zone; thick solid lines 
in Figure 1) after the galaxy distribution is smoothed in the
S-zone. There are 605 galaxies with $z\le 2$ and $AB(8140)\le 28.0$
in the G-zone.
To see the possible evolution effects we divide the HDFN sample into
three overlapping redshift slice subsamples, each of which contains about 
300 galaxies in the G-zone. Table 1 lists the definitions
of these subsamples together with the total sample.
The middle slice overlaps with the first and the third ones.

The photometric redshift data in the HDF South (HDFS) field has been
obtained from a web page\altaffilmark{1} (Lanzetta et al. 1999; 
Yahata et al. 2000).  This catalog is
close to complete down to $AB(8140)\le 28.0$ (Fern\'andez-Soto 2000).
The original catalog contains 1275 redshifts. The magnitude limit of
$AB(8140)= 28.0$ and redshift limit of $z=3$ leave us with 727 galaxies,
and 614(530) of these galaxies are within the S(G)-zones, respectively.
This whole HDFS sample is further divided into three redshift space
slices with roughly 265 galaxies each as described in Table 1.
Figure 2 shows the HDFS galaxies satisfying the magnitude and redshift
limits, and the boundaries defining the S- and G-zones.

\altaffiltext{1}{http://www.ess.sunysb.edu/astro/hdfs}

\subsection{Results}

To measure the genus the discrete distribution of galaxies
must be smoothed by an appropriate filter. 
We first make a mask which has the value 1 within the S-zone and 0 outside. 
We ignore the small regions contaminated
by stars and relatively big galaxies in the HDFs because
taking into account those regions to the mask
turns out to have negligible effects on the genus results because
our smoothing lengths are large compared to them.
We first smooth the mask over a smoothing length 
using the Gaussian filter. At the same time the distribution of
the HDF galaxies in the S-zone are smoothed over the same length
and divided by the smoothed mask to yield the smooth galaxy density 
field (Melott et al. 1989).
Then the genus is measured from the smoothed density array only within
the G-zone.
We use the CONTOUR2D code (Weinberg 1988) to measure the genus.
The code has been modified so that the genus at each threshold
level is an average over three genus values with the threshold levels
$\nu$ shifted by 0 and $\pm 0.03$.

We have chosen the Gaussian smoothing radius $R_G$ as the half-width at
half maximum of the Gaussian function, 
and set $R_G = {\bar d}/ \sqrt{2{\rm ln}\; 2}=0.849 {\bar d}$. This is 
slightly smaller than the mean separation ${\bar d}$, but larger than the
`e-folding smoothing length' $\lambda_e = {\bar d}/\sqrt{2}=0.707 {\bar d}$
used in some studies.

Figure 3 and 4 show the genus curves (filled dots) for our HDFN and 
HDFS samples and their best fitting Gaussian genus curves (solid curves), 
respectively. Error bars are estimated from 20 bootstrap resamplings
of the galaxies.  Small open circles are the genus curves averaged over 100 
realizations of Poisson distributions with the same number of galaxies.
In Table 2 we summarize the genus-related statistics $A, \Delta \nu$,
and $\Delta g$ measured from each HDF subsample. 
The 68\% uncertainty limits are again estimated from the genus-related
statistics measured from 20 bootstrap resamplings of galaxies.
The mean amplitude of the
genus curves from 100 Poisson realizations of the distribution of
the mock HDF galaxies in each sample
is also included.

\section{Discussion}

The genus curves of the HDF samples shown in Fig. 3 and 4 and 
their statistics listed in Table 2 indicate that the 
distribution of the HDF galaxies is consistent with a Gaussian
random phase distribution
because most subsamples have shift and asymmetry parameters
consistent with zero. The only statistically significant
behaviour of the genus curves is their lower amplitudes compared to
the corresponding Poisson distributions.
In the north samples this coherence of the galaxy distribution
is mainly caused by the shallow slice with $0\le z<1.1$.
The fact that the genus curve has an amplitude significantly
lower than the Poisson one indicates that the smoothed distribution of the
HDF galaxies does have a real signal.

Therefore, even though the our HDF samples are radial projections
of galaxies in very long thin rods, the galaxy distribution in each sample
is not merely a projection of statistically independent galaxies,
but maintains a finite clustering signal.
To check if this should be the case, we need to know the angular 
covariance function (CF) of galaxies at high redshifts.
Gott and Turner (1979) have found that the angular CF of
galaxies at the present epoch continues inward without any break
with a slope of $0.8$ (i.e. $w(\theta)\propto \theta^{-0.8}$) down to
the smallest scale measured, a comoving scale of 0.0033 $h^{-1}$ Mpc.
For comparison at the median redshift $z=1.1$ of the HDFN (for an
Einstein-de Sitter cosmology) the smoothing lengths $R_G = 3.8''$
and $5.4''$ correspond to comoving scales of $0.034 h^{-1}$ Mpc and
$0.049 h^{-1}$ Mpc, respectively (or over an order of magnitude larger).
Assuming that the CF is constant in the comoving space
(The CF does not grow greatly from $z=1.1$ to the present in the
flat lambda model with $\Omega_m = 1/3$ and $\Omega_{\Lambda}=2/3$ 
popular today, for example.), from
Gott and Turner's present epoch CF from the Zwicky catalog
$w(\theta)\approx 17.3(\theta/1')^{-0.8}$
we estimate
$$w(\theta) = 0.494 (\theta/1.7'')^{-0.8}=(\theta/0.70'')^{-0.8}, \eqno (9)$$
as the depth of the sample changes from $D_* = 53 h^{-1}$ Mpc for the Zwicky
catalog to $D_* = 1860 h^{-1}$ Mpc for the HDFN sample.
Then the average fractional excess number counts 
of galaxies within a smoothing area around a galaxy is approximately
$$\langle{\delta N\over N} \rangle = 
  {1\over \pi R_G^2}\int_0^{R_G} w(\theta) 2\pi \theta d\theta
  =  {5\over 3} (R_G/0.70'')^{-0.8}. \eqno (10)$$
Since $R_G = 3.8''$ for the HDFN, $\langle{\delta N\over N}\rangle
=0.43$. For comparison, a Poisson distribution would have
a RMS fluctuation of $\langle{\delta N\over N}\rangle
= 1/\sqrt{N} =  (R_G/{\bar d})^{-1}/\sqrt{\pi} = 0.66$ where we used
$N = \pi R_G^2/{\bar d}^2$ given the mean galaxy separation ${\bar d}$.
Therefore, we expect the signal-to-noise ratio in the HDFN sample to be
about 0.65, so there still remains some detectable 
physical clustering of galaxies
in the field.
A lower limit on the signal-to-noise ratio could be established by
assuming that clusters present at $z=1.1$ remained at the same
physical size to the present, representing a growth of the amplitude
CF in comoving coordinates proportional to $a^{2-0.8}=a^{1.2}$ 
where $a=1/(1+z)$.
That would give a $S/N=0.27$.

We may also estimate the signal-to-noise ratio in the smoothed maps directly 
from the topology statistics. The amplitude of the genus curve is
proportional to $F(n)$ from equation (5). The noise contribution
from a Poisson distribution of galaxies corresponds to
a power-law index $n=0$ and $F(n)=1$. If the signal is 
characterized by an angular CF with $w(\theta )\propto \theta^{-0.8}$.
This corresponds to a power-law index $n=-1.2$ and $F(n)=0.4$.
Thus if we were observing pure signal we would expect
$A=0.4 A_{\rm Poisson}$. If we were observing pure noise we would expect
$A=A_{\rm Poisson}$. We actually observe for the HDFN a value of
$A=0.77 A_{\rm Poisson}$ suggesting, by linear interpolation,
a value of $S/N = 0.62$. Repeating this calculation for the HDFS where
we observe $A=0.86A_{\rm Poisson}$ gives a $S/N=0.30$,
both being consistent with the back-of-the-envelope calculations given
in the previous paragraph.

We have also found that the shift parameter in Table 2 is consistent with
zero shift. Actually the HDFN-1 and HDFS samples show significant
bubble ($\Delta\nu>0$) and meat-ball shifts, respectively.
But the shift averaged over different subsamples is consistent
with zero.
The asymmetry parameter is also consistent with zero. Even though
the HDFN-3 sample has a significant excess of the genus curve amplitude
in the positive thresholds (more clusters than voids at the same
volume fraction), the mean asymmetry parameter for all
subsamples is still within
one standard deviation from zero.

\acknowledgments

This work was supported by the KOSEF grant (1999-2-113-001-5) and
the NSF grant AST-9900772.
CBP would like to thank Canadian Institute for
Theoretical Astrophysics for the hospitality during this work.
The authors thank Dr. Michael Vogeley for helpful comments 
and Dr. Alberto Fern\'andez-Soto for sending us unpublished results 
on the HDFS photometric redshift data.

\clearpage


\begin{figure}
\plotone{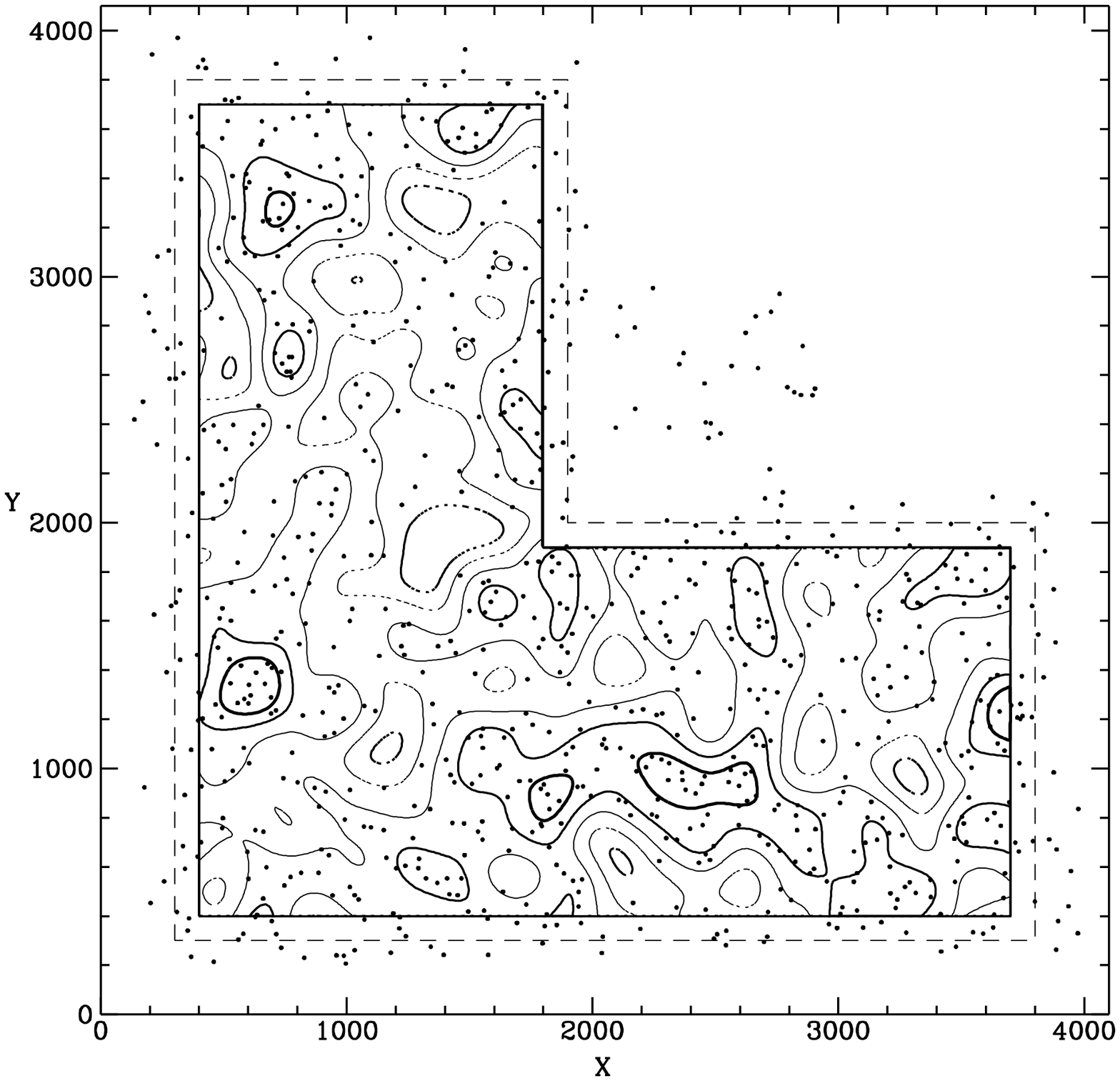}
\caption{
The 820 galaxies (dotts) with $0\le z \le 2.0$ in the HDF North.
The dashed lines mark the region where the galaxy distribution is smoothed,
and the solid lines show the borders of the region where the genus is 
measured. Contour lines represent the $\nu_{A}=+2$ (thick solid line), 
$+1$ (solid), $0$ (thin solid), $-1$ (light dotted), and $-2$ (heavy
dotted) iso-density contours of the galaxy number density smoothed over
$3.8''$ or 95 pixels. 
Coordinates are the pixel numbers in the HDF mosaic image.
}
\end{figure}

\begin{figure}
\plotone{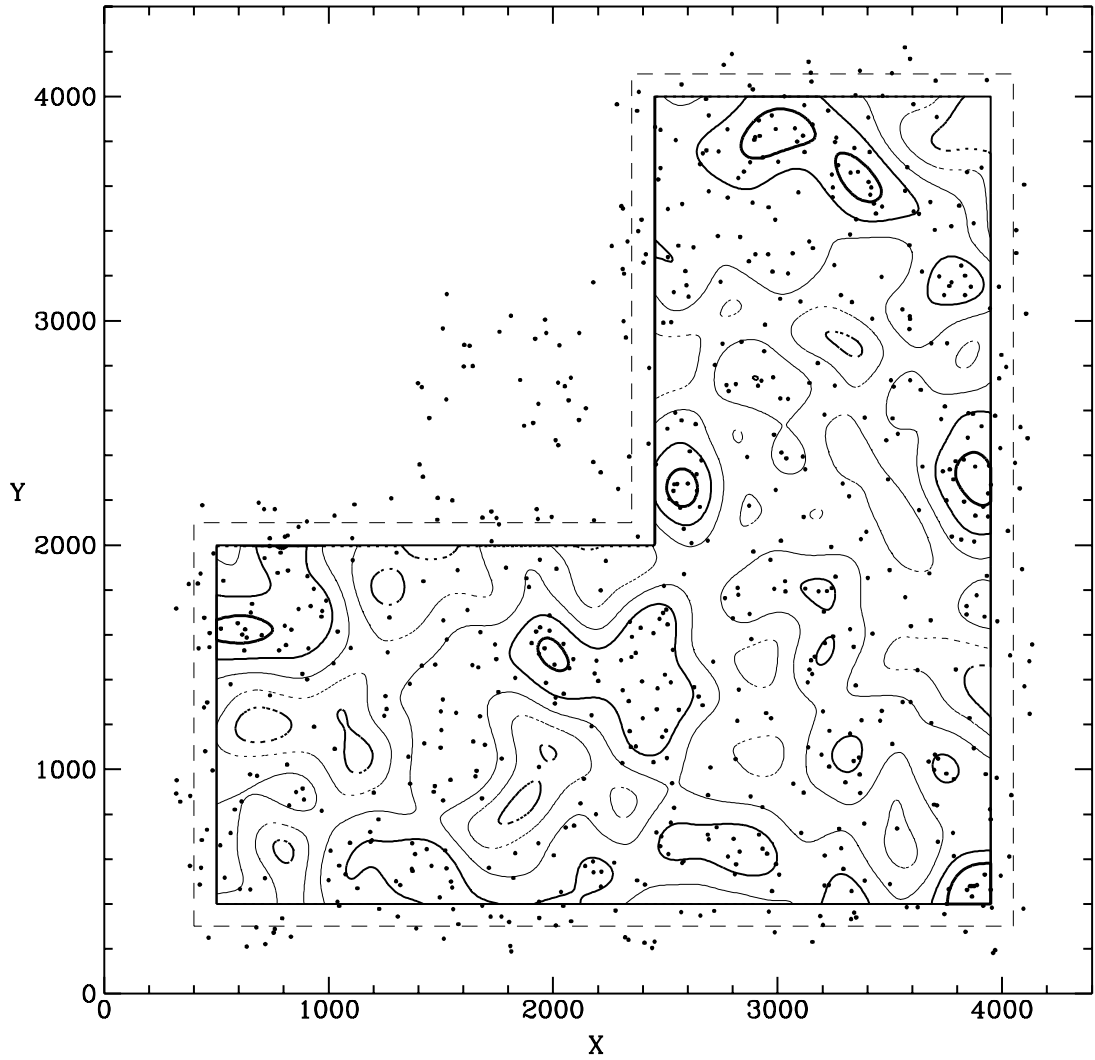}
\caption{
The 727 galaxies (dotts) with  $0\le z \le 3.0$ in the HDF South.
The mosaic image is rotated by $4.76^{\circ}$ clockwise.
Superposed are the iso-density contours of the galaxy distribution smoothed
over $4.4''$ or 110 pixels.
}
\end{figure}
\begin{figure}
\plotone{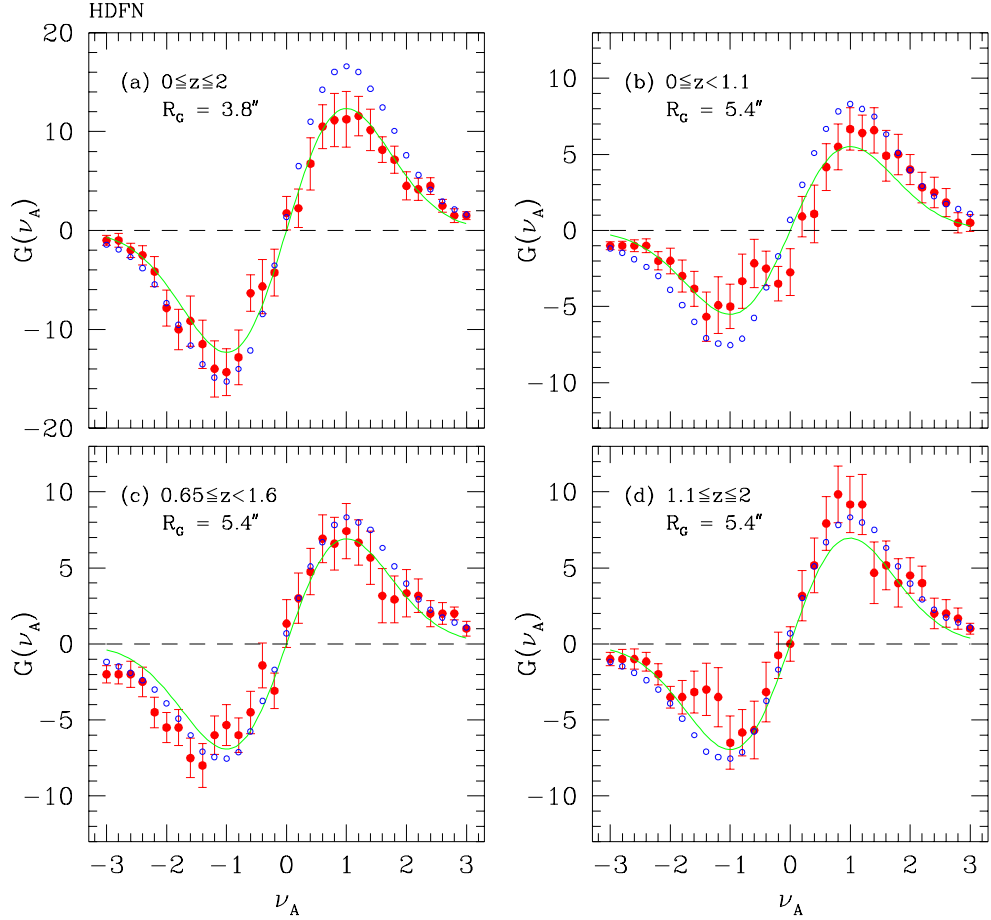}
\caption{
The genus curves (filled dotts) of the galaxy distributions of the HDF 
North subsamples. The solid lines are the Gaussian genus curves
best fit to the observed data. The open circles are the average 
genus curve of 100 Poisson distributions with the same number of
points as the number of galaxies in each HDF subsample.
}
\end{figure}

\begin{figure}
\plotone{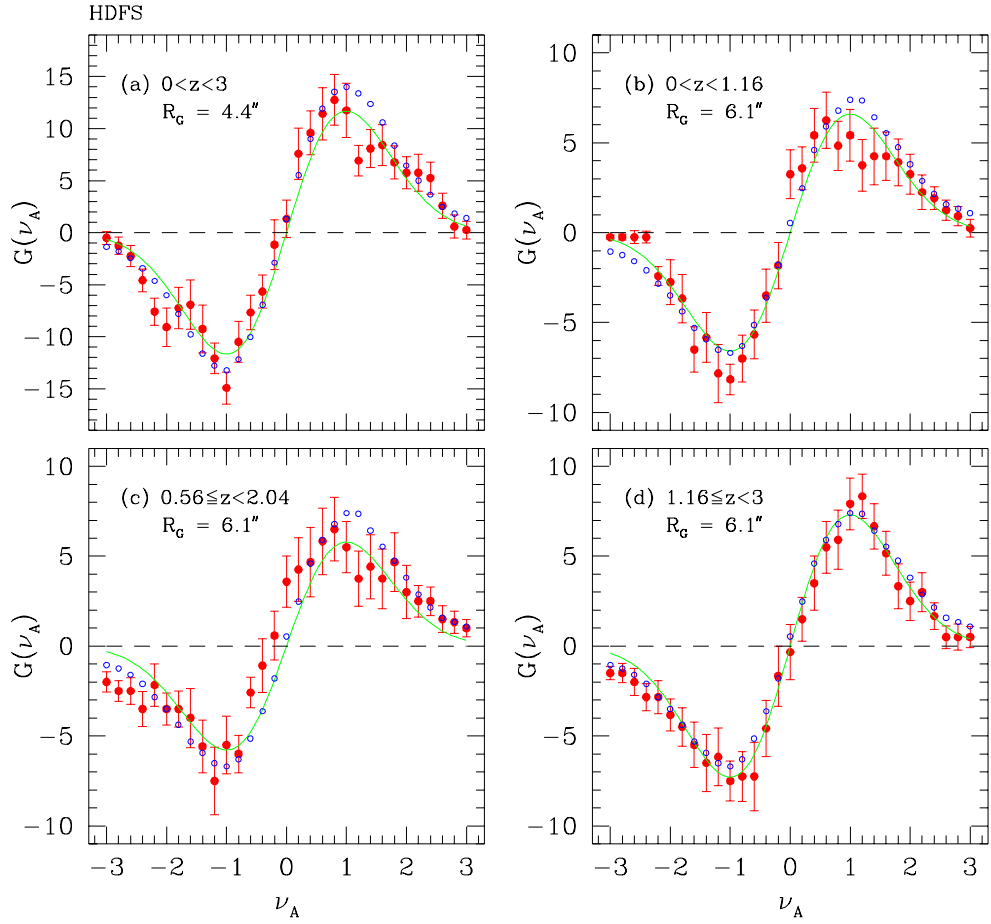}
\caption{
Same as Fig. 3 but for the HDF South subsamples.
}
\end{figure}

\clearpage

\begin{deluxetable}{p{2.0in}cccc}
\tablewidth{0pt}
\tablecaption{
\label{tbl:stars}
The HDF subsamples
}
\tablehead{
\colhead{Samples} &
\colhead{$z$} &
\colhead{$z_{\rm med}$} &
\colhead{$N_g$(S/G-zone)} &
\colhead{$R_G$}
}
\startdata
HDFN  \dotfill  & $0.00\le z\le 2.00$ & 1.10 & 714/605 & $3.8''$ \\
HDFN-1 \dotfill & $0.00\le z <  1.10$ & 0.65 & 368/307 & $5.4''$ \\
HDFN-2 \dotfill & $0.65\le z <  1.60$ & 1.10 & 365/305 & $5.4''$ \\
HDFN-3 \dotfill & $1.10\le z\le 2.00$ & 1.60 & 346/298 & $5.4''$ \\
& & & & \\
HDFS  \dotfill  & $0.00\le z\le 3.00$ & 1.16 & 614/530 & $4.4''$ \\
HDFS-1 \dotfill & $0.00\le z<   1.16$ & 0.56 & 300/204 & $6.1''$ \\
HDFS-2 \dotfill & $0.65\le z<   2.04$ & 1.16 & 310/364 & $6.1''$ \\
HDFS-3 \dotfill & $1.16\le z\le 3.00$ & 2.04 & 314/266 & $6.1''$ \\
\enddata
\end{deluxetable}

\clearpage

\begin{deluxetable}{p{2.0in}cccc}
\tablewidth{0pt}
\tablecaption{
\label{tbl:stars}
Genus-related statistics for the HDF subsamples
}
\tablehead{
\colhead{Samples} &
\colhead{$A$} &
\colhead{${\bar A}_{\rm Poisson}$} &
\colhead{$\Delta\nu-\Delta\nu_{\rm Poisson}$} &
\colhead{$\Delta g - \Delta g_{\rm Poisson}$}
}
\startdata
HDFN   \dotfill & $20.3\pm 2.3$ & 26.5 & $-0.01\pm 0.08$ & $-0.22\pm 0.17$  \\
HDFN-1 \dotfill & $ 9.1\pm 1.5$ & 13.2 & $+0.30\pm 0.10$ & $+0.21\pm 0.21$  \\
HDFN-2 \dotfill & $11.4\pm 1.5$ & 13.2 & $-0.01\pm 0.11$ & $-0.14\pm 0.19$  \\
HDFN-3 \dotfill & $11.5\pm 2.7$ & 13.2 & $+0.03\pm 0.07$ & $+0.35\pm 0.13$  \\
& & & & \\
HDFS   \dotfill & $19.2\pm 1.9$ & 22.4 & $-0.15\pm 0.07$ & $-0.12\pm 0.14$  \\
HDFS-1 \dotfill & $10.6\pm 0.9$ & 11.7 & $-0.17\pm 0.10$ & $-0.31\pm 0.16$  \\
HDFS-2 \dotfill & $ 7.2\pm 1.2$ & 11.7 & $-0.25\pm 0.15$ & $-0.24\pm 0.28$  \\
HDFS-3 \dotfill & $12.0\pm 1.1$ & 11.7 & $+0.03\pm 0.11$ & $-0.19\pm 0.16$  \\
\enddata
\end{deluxetable}

\clearpage

\end{document}